\documentstyle[11pt,epsf]{article}
\newcommand{\bm}[1]{\mbox{\boldmath$#1$}}

\newcommand{\Kc}{K_{c}}
\newcommand{\dudk}{\frac{dU_{L}}{dK}}
\newcommand{\Kchi}{K_{\chi^c_{max}}}
\newcommand{\abs}[1]{\left| #1 \right|}
\newcommand{\bra}{\left\langle}
\newcommand{\ket}{\right\rangle}
\newcommand{\e}{\mbox{e}}
\newcommand{\sx}[1]{\vec{s}_{\bm #1}}
\newcommand{\sxd}[1]{\vec{s'}_{\bm #1}}

\title{\begin{flushright}
\vspace{-2.0cm}
{\normalsize
 UTHEP-284\\
 September 1994 \\}
\end{flushright}
\vspace*{2.0cm}
\bf Critical exponents \\of a three dimensional 
O(4) spin model}

\author{ 
{\bf K. Kanaya and S. Kaya}\\
{\small   \it Institute of Physics,  University of 
Tsukuba, Ibaraki 305, Japan}
}
\date{}
\begin{document}

\maketitle

\begin{center}
{\bf Abstract\\}
\end{center}

By Monte Carlo simulation
we study the critical exponents governing the transition of
the three-dimensional classical O(4) Heisenberg model, 
which is considered to be in the same universality class as
the finite-temperature QCD with massless two flavors.
We use the single cluster algorithm and the histogram reweighting 
technique to obtain observables at the critical temperature.
After estimating an accurate value of the inverse critical 
temperature 
$\Kc=0.9360(1)$, we make non-perturbative estimates for various 
critical exponents by finite-size scaling analysis.
They are in excellent agreement with those obtained with the 
$4-\epsilon$ expansion method with errors reduced to about 
halves of them.

\vfill \eject

\newpage

\section{Introduction}
Finite temperature chiral phase transition of QCD is very 
important in the study of phase transitions in the early
Universe and in the investigation of heavy ion collisions 
at high energy.
At present, this transition is studied mainly using 
the Monte Carlo method
on the lattice.
Pisarski and Wilczek \cite{wilczek,wilczek2} suggested that 
QCD with massless two flavors, 
which is considered to be an approximation of the real world, 
belongs to the same universality class as 
three-dimensional four-component Heisenberg models, 
if the finite temperature chiral transition of $N_f=2$ QCD is 
second order. 
Then, the chiral transition of $N_{f}=2$ QCD 
has the same critical exponents as the 3d O(4) Heisenberg model.

Simulations of lattice QCD for $N_{f}=2$ suggest that the chiral 
transition is a second order transition for staggered fermions 
\cite{fukugita} and for Wilson fermions \cite{tsukuba}. 
The study towards a precise measurement of the critical exponents 
of $N_{f}=2$ QCD has just begun~\cite{karsch}. 
In the verification that 
the O(4) Heisenberg model belongs to the same universality class, 
there is a problem that both Wilson fermions and staggered 
fermions 
on the lattice do not have the full chiral symmetry --- 
which is expected to restore only in the continuum limit.
Conversely, however, we could consider that, assuming the 
universality,
the chiral symmetry is restored on the lattice sufficiently
when the exponents agree with those of the 3d O(4) Heisenberg 
model.

Therefore an accurate calculation of the critical exponents of
the 3d O(4) Heisenberg model is quite important.
For this model the best estimation of critical exponents has 
been made with the $4-\epsilon$ expansion method up to seven 
loops~\cite{baker}.

In this work we simulate the 3d O(4) Heisenberg 
model by the Monte Carlo method
and make a non-perturbative estimation of several critical 
exponents.
We use the single cluster Monte Carlo update algorithm
which recently has been used for the simulation of spin systems:
Wolff formulated this algorithm by modifying the multiple-cluster 
algorithm by Swendsen and Wang\cite{wang1} and 
applied it to continuous spin models\cite{wolff1,wolff3}.
Recent applications of the multiple and single cluster algorithms
to two and three dimensional spin models 
have demonstrated their advantage in the computation time to 
the usual local update algorithms.
Among global algorithms,
the single-cluster algorithm is shown to be superior to the
multiple-cluster algorithm  
for three-dimensional spin models\cite{wolff2,multi}.
Therefore, we apply the single-cluster algorithm in this study. 

In section \ref{sec:2} the model and the method of simulation 
are described.
In section \ref{sec:3} 
we estimate the transition point from the crossing point of 
the Binder cumulant and compute the critical exponents
at the transition point making use of the histogram reweighting 
technique.
We also check the consistency of the results by independent 
measurements 
of the critical temperature and several exponents. 
We then compare our exponents with those of the $4-\epsilon$ 
expansion method.
Our conclusion is given in section \ref{sec:4}.

\section{The model and the method}
\label{sec:2}

The partition function $Z$ and the energy $E$
of the 3d O(4) Heisenberg model
are defined by
\begin{eqnarray}
Z &=& \prod_{\bm x}\int\left[d\vec s\right]\e^{\left(-KE\right)},
\nonumber \\
E &=& 
\sum_{\bm x,\hat{\bm{ \imath}}}\left\{1-\vec s(\bm x)\cdot \vec 
s(\bm{x}+\hat{\bm \imath})\right\},
\end{eqnarray}
where $K$ is the inverse temperature and 
$\vec s(\bm x)$ is a four-dimensional unit spin at the lattice 
site $\bm x$.
$\hat{\bm{ \imath}}$'s are the unit steps in three coordinate 
directions.  
We use three-dimensional simple cubic lattices 
with the volume $V=L^3$ with $L=8$, 10, 12, 14, 16, 24, and 32, 
and employed periodic boundary conditions.

We chose two simulation points for each $L$ except for $L=10$: 
One is $K=0.935$ which is a rough estimate for the 
transition point by our preparatory simulation.
Another simulation point is chosen for each $L$ at the maximum 
point of the 
susceptibility estimated by a preparatory simulation. 
Our simulation parameters are compiled in Table \ref{tbl:1}.
We use the data at $K=0.935$ for the calculation of the 
transition point 
as well as the analyses of finite-size scaling
with the histogram reweighting technique
and use the data at the maximum of the susceptibility 
for a check of the consistency of our results.

The magnetization and the energy are measured every 10 sweeps
and stored on the disk. 
We define one sweep by one cluster update by the single-cluster 
algorithm
explained in the following section.
Several million sweeps are performed for each simulation point. 
From the autocorrelation time we measured (see the following 
section) 
this corresponds to about one hundred thousand independent data 
for 
each point, as compiled in Table \ref{tbl:1}. 
We estimate errors by the jackknife procedure.
We study the bin-size dependence of erros and choose a suffiently
large bin-size such that errors become stable.
The resulting bin-sizes are consistent with 
the values of auto-correlation time estimated independently.
All the jobs takes 23 hours with HITAC S820/80.

We use the histogram reweighting method\cite{swendsen}
to calculate the observables in a region of $K$ around the 
simulated point
$K_{simu}$.
The region of $K$ in which the histogram reweighting method is 
applicable 
can be determined by the magnitude of the shift of energy value: 
If the peak position of a reweighted energy distribution, 
$E_{peak}(K)$, locates 
away from the peak position of the original distribution, 
$E_{peak}(K_{simu})$, 
then the statistical errors for averages computed with the 
reweighted 
distribution become large correspondingly. 
Limited statistics near the tails of measured histograms also 
leads to a danger of large under-estimation of the errors there. 
We study the effect of reweighting and observe that, with our 
statistics, 
many errors for the observables we study become rapidly large and 
the histogram becomes rapidly notched when $K$ gets outside the 
region 
where the height of the original energy histogram at 
$E_{peak}(K)$ is larger
than one third of the peak height. 
Although several computed errors, such as the error for 
the Binder cumulant 
discussed below, sometimes remain small even outside this range, 
we find that 
the result is not consistent with the result of a direct 
simulation there. 
We therefore limit ourselves to apply the histogram reweighting 
method 
only up to the point where the height of the original energy 
histogram 
at $E_{peak}(K)$ decreases to a third of the peak height. 
Similar criterion is used also in Ref.~\cite{holm}.

\subsection{Algorithm}
We use the single-cluster algorithm formulated by 
Wolff\cite{wolff1}.
This is a global update algorithm whose advantage is that
the autocorrelation time and the dynamical exponent are both 
much smaller
than those of the local update algorithm as discussed below. 

The autocorrelation function $A(k)$ is defined by
\begin{eqnarray}
A(k) &=& \frac{\rho(k)}{\rho(0)}\;, \nonumber\\
\rho(k) &=& \bra O_{i}O_{i+k}\ket - \bra O_{i} \ket ^2,
\end{eqnarray}
with $O_{i}$ being the i-th measurement of an observable $O$.
The autocorrelation time $\tau$ which is given by 
integrating the autocorrelation function
\begin{equation}
\tau = \sum _{k=1} ^{\infty}A(k) 
\label{eqn:tau}
\end{equation}
diverges in the critical region as $\displaystyle \tau \propto 
\xi^{z}$, 
where $\xi$ is the correlation length. 
The exponent $z$ is called as dynamical exponent.
On finite lattices in the critical region, 
$\xi$ is replaced by the lattice length $L$: 
\begin{equation}
\tau \propto L^{z}\;.
\label{eqn:taul}
\end{equation}
This lattice size dependence of $\tau$ is the origin of the 
``critical slowing down'' which makes difficult to get a high 
effective 
statistics in the critical region on large lattices. 
We should use an algorithm with a small dynamical exponent.
It is known that the local update algorithms 
such as the Metropolis algorithm have $ z \sim 2$ independent 
of the model 
and the details of the update algorithm.
For example, $z=1.94(6)$ is obtained for the 3d O(3) Heisenberg 
model 
with Metropolis algorithm\cite{peczak1}. 
Use of a global update algorithm is required to get a smaller $z$.
It is reported in Refs.\cite{wolff2,holm,wang3} that $z$ with the 
single-cluster algorithm for the 3d Ising model is about 0.2 and 
that for the 3d O(3) Heisenberg model is about 0. 
As presented in the next section, our result of $z$ for 
the 3d O(4) Heisenberg model is also consistent with 0.

The single-cluster update 
for the O($n$) Heisenberg model is as follows 
\cite{wolff1,evertz}:
\begin{enumerate}
\item A unit vector in the O($n$) space, $\vec r$,
is chosen with a random direction. 
\item A starting site of a cluster, $\bm x_{0}$, is chosen 
at random 
and is included  in the cluster.
\item  For a link on the surface of the cluster,
$\partial_C =\{(\bm x,\bm y)|\bm x\in C\:,\; \bm y\not\in C)\}$,
$\bm y$ is included in the cluster with the probability
\begin{equation}
	P(\vec s _{\bm x},\vec s _{\bm y})
	= 1-\exp [{\min \{ 0,-2K(\vec r\cdot \vec s_{\bm x})
	(\vec r\cdot \vec s_{\bm y}) \} ]\;.
}
\end{equation}
\item 
The process (3) is repeated until the growth of the cluster stops.
\item All spins in the cluster is flipped with regards to
the surface perpendicular to $\vec r$:
\begin{equation}
\sxd x =R(\vec r)\sx x=\sx x - 2(\sx x \cdot \vec r)\vec r\;.
\end{equation}
\end{enumerate}

In order to test the efficiency of the algorithm 
and to test our program code for the single-cluster update,
we simulate the 3d Ising and the 3d O(3) Heisenberg model. 
Our results are completely consistent with 
Refs.\cite{wolff2,holm},
including the results for susceptibility, dynamical exponent, 
and critical
exponents.

\section{Results}
\label{sec:3}

\subsection{Autocorrelation time and energy distribution}
Our results for the autocorrelation time are compiled in Table 
\ref{tbl:1}.
The autocorrelation time $\tau_m$ in terms of Metropolis unit
stays almost constant or rather decreases with 
the increase of the lattice size.
This implies that the dynamical exponent is 0 or slightly 
smaller than 0. 
Similar result is obtained also for the O(3) Heisenberg 
model\cite{holm}.

\begin{figure}[p]
\epsfxsize=12cm \epsfbox{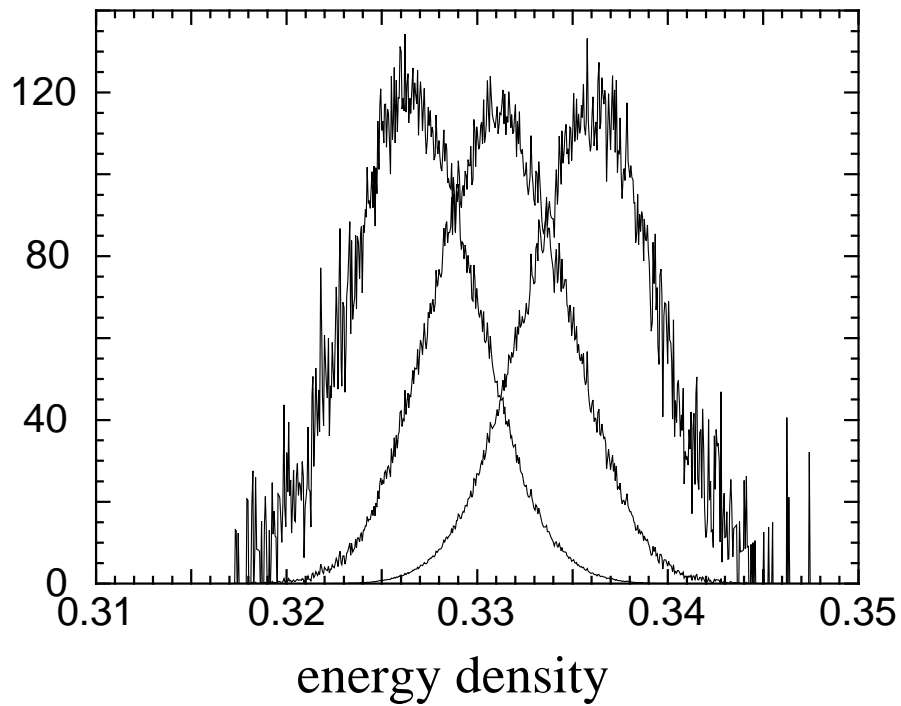}
\caption{
Histogram of the energy density 
for $L$=32 near the critical coupling.
Each distribution is normalized to unit area. 
Histograms are for $K=0.929$, 0.935, and 0.939 
from the left to the right, respectively. 
The histograms for $K=0.929$ and 0.939 are obtained by 
reweighting the 
measured histogram at $K$=0.935. 
}
\label{fig:1}
\end{figure}

The measured energy distribution shown in Fig.~\ref{fig:1}
is a Gaussian type with single peak.
The continuous shift of the distribution with the temperature 
over the 
expected critical region is consistent with a second order 
phase transition 
in accord with the results of the $4-\epsilon$ expansion method. 
Final confirmation of the order of the transition is done with 
the values of the critical exponents discussed below. 

\subsection{Critical temperature}
\begin{figure}[p]
\epsfxsize=12cm \epsfbox{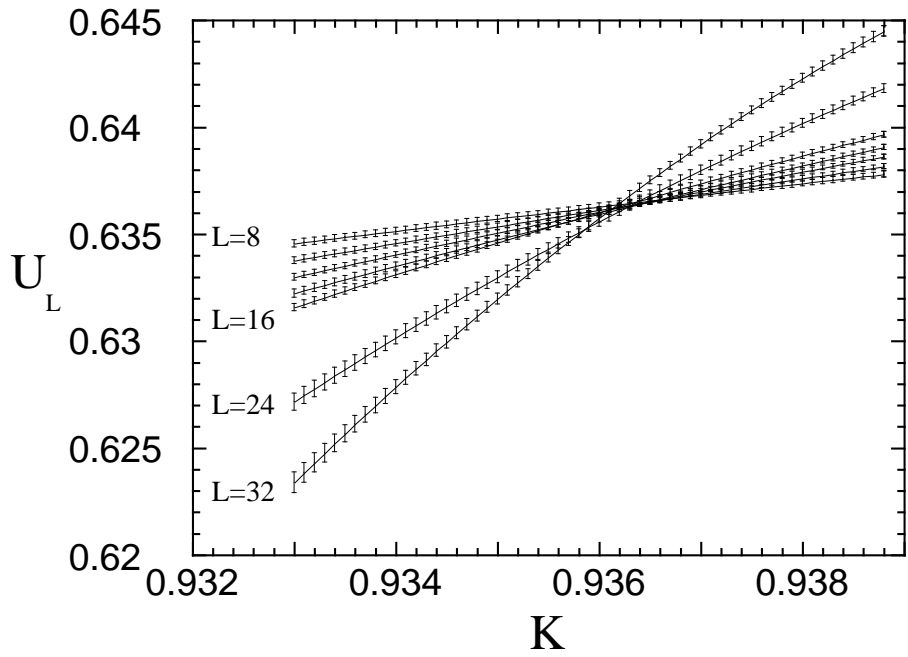}
\caption{
The Binder cumulant $U_L$ as a function of the inverse 
temperature $K$
for $L=8$, 10, 12, 14, 16, 24, and 32.
$U_L(K)$ is computed with the histogram reweighting method 
using the 
data at $K=0.935$.
}
\label{fig:2}
\end{figure}

Accurate calculation of critical exponents requires a precise 
determination
of the inverse critical temperature $K_c$.
An efficient method to determine $K_c$ for a second order 
transition is to measure the Binder cumulant 
\cite{binder} 
for various system size and to locate the cross point 
in the space of $K$.
On sufficiently large lattices where subleading corrections from 
the 
finite lattice size $L$ are ignored, 
the Binder cumulant $U_{L}(K)$ defined by 
\begin{eqnarray}
U_{L}(K) &=& 1-\frac{1}{3}\frac{\bra m^4 \ket}{{\bra m^2 \ket}^2}
 \nonumber \\
\vec m &=& \frac{1}{V}\sum_{\bm x} \vec s({\bm x})\;,
\end{eqnarray}
becomes independent of $L$ at the transition point $K_c$ 
\cite{binder}:
\begin{equation}
\frac{U_{L'}(\Kc)}{U_{L}(\Kc)}=1,
\label{eqn:4}
\end{equation}
and the slope of $U_{L}(K)$ in $K$ at $K_c$ increases as $L$ 
becomes large.
In Fig.\ref{fig:2} are shown our results
of the Binder cumulant near the crossing point. 
The values for $U_{L}(K)$ are 
obtained with the histogram method using the data at $K$=0.935.

The deviation from the relation (\ref{eqn:4}) observed 
in Fig.\ref{fig:2}
can be explained by the finite-size confluent corrections. 
The leading $L'/L$-dependence in the deviation of the crossing 
point $K^*$ 
from the critical point $K_c$ is estimated by Binder\cite{binder}
 as
\begin{equation}
\frac{1}{K_c}-\frac{1}{K^*} \propto \frac{1}{\ln b}
\end{equation}
where $b=L'/L$.

\begin{figure}[p]
\epsfxsize=12cm \epsfbox{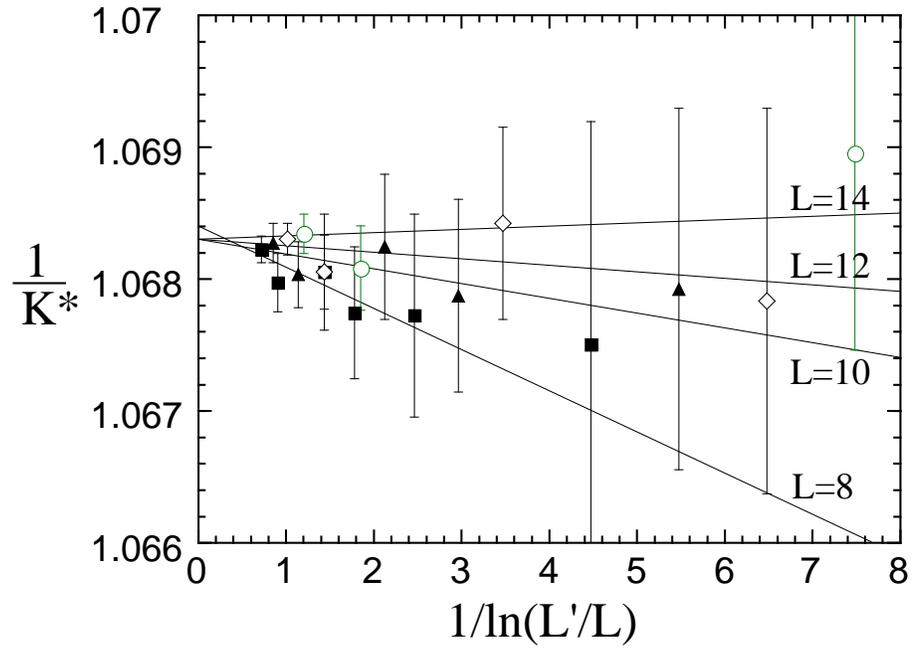}
\caption{
The crossing point of Binder cumulants $U_L(K)$ and $U_{L'}(K)$ 
for 
$L=8$(squares), 10(triangles), 12(diamonds), and 14(circles)
with different $L'$.  
Solid lines correspond to linear least-square fits for each $L$.
The critical coupling is estimated as $K_c=0.9360(1)$ by 
extrapolating these lines to the limit $1/\ln (L'/L)=0$. 
}
\label{fig:3}
\end{figure}

We plot $(1/{\ln b},1/{K^*})$ for $L=8$, 10, 12, and 14 
in Fig.~\ref{fig:3}.
The errors for $1/{K^*}$ are computed from the jackknife 
errors for 
$U_{L}(K)$.
The solid lines in Fig.~\ref{fig:3} represent the results 
of a linear 
least-square fit for each $L$.
We find that the correction with $1/{\ln b}$ is smaller than
that of 3d O(3) Heisenberg model \cite{holm}.
The extrapolation of $1/K^*$ to the point $1/{\ln b}=0$ 
for each $L$ 
gives the values for $1/K_c(L)$:
$1/{K_c}(8)=1.06841(21),$ $1/{K_c}(10)=1.06832(26),$ $
1/{K_c}(12)=1.06833(26),$ $1/{K_c}(14)=1.06826(34)$.
All of them are consistent with each other.
The mean value of these results is $1/{K_c} = 1.06835(13)$.
A similar fit for all $L$ with a common parameter $1/K_c$
gives the value $1/{K_c} = 1.06836(14)$ which completely agrees 
with the mean value.
We quote hereafter 
\begin{eqnarray}
\frac{1}{K_c}&=&1.06835(13),\\
\Kc&=&0.9360(1).
\end{eqnarray}

\subsection{The critical exponent $\nu$}
\begin{figure}[p]
\epsfxsize=12cm \epsfbox{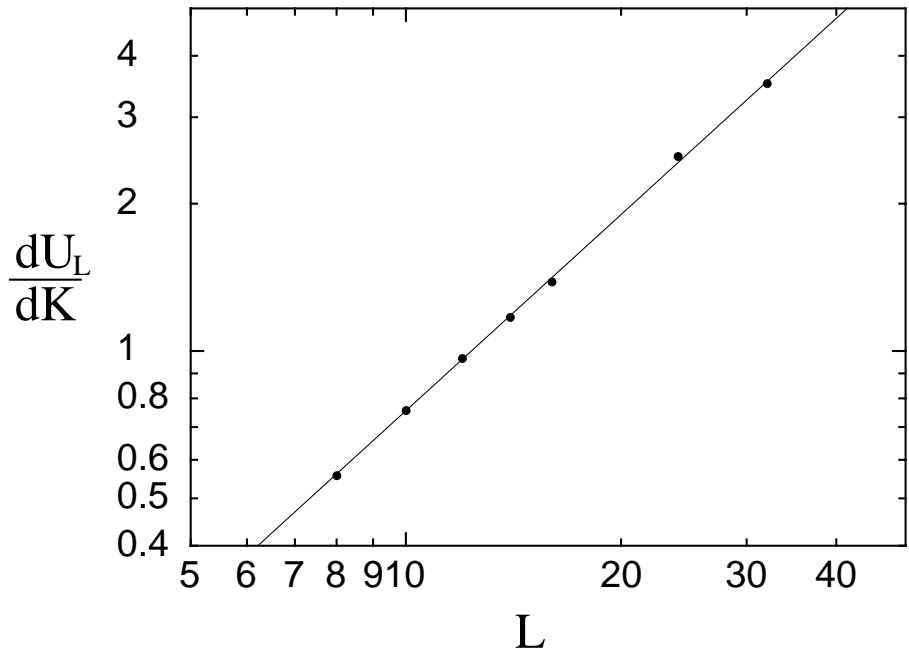}
\caption{
Scaling of the slope $dU_L/dK$ of the Binder cumulant
at $K_c=0.9360(1)$ as a function of the lattice size $L$. 
The slope of the solid line given by a linear least-square 
fit leads to 
an estimate of the critical exponent $1/\nu=1.337(16)$.
The jackknife errors for $dU_L/dK$ are smaller than the size 
of symbols.
}
\label{fig:4}
\end{figure}

The slope for $\displaystyle\left. \dudk \right| _{K=\Kc}$ 
is known to scale with a critical exponent $\nu$ as
\cite{binder}
\begin{equation}
\left. \dudk \right| _{K=\Kc} \sim L^{1/\nu}\;.\label{eqn:scale1}
\end{equation}
Using the relation
\begin{equation}
\dudk =(1-U_{L})\left\{\bra E \ket - 2\frac{\bra m^2 E \ket}
{\bra m^2 \ket}
                    +\frac{\bra m^4 E \ket}{\bra m^4 \ket}
             \right\}
\end{equation}
we calculate $\displaystyle\left. \dudk \right| _{K=\Kc}$ 
at the estimated $\Kc=0.9360$.
In Fig \ref{fig:4}, we plot $\displaystyle \dudk$ as a function 
of $L$. 
From the slope of the solid line in this logarithmic plot 
we find 
\begin{eqnarray}
\frac{1}{\nu} &=& 1.337(16),\\
\nu &=& 0.7479(90).
\end{eqnarray}
by a least-square fit. 
We repeat the analysis by varying $\Kc$ within our estimated 
error, 
0.9360(1), and find that the results for $\nu$ are completely 
consistent with the result given here. 

The scaling relation (\ref{eqn:scale1}) requires a sufficiently 
large $L$ 
to ignore the subleading corrections. 
In order to test if our values of $L$ is large enough, we repeat 
the fits excluding the data for the smallest size $L=8$, and
for $L=8$ and $L=10$. 
We obtain $1/{\nu}$($L=8$ excluded)=1.344(36) and $1/{\nu}$($L=8$ 
and 10 excluded)=1.333(51), respectively. 
Because these results are completely consistent with $1/{\nu}$ 
with all data, 
we conclude that $L=8$ is sufficiently large to extract scaling 
properties.

\subsection{The result for $\displaystyle \frac{\beta}{\nu}$}
\begin{figure}[p]
\epsfxsize=12cm \epsfbox{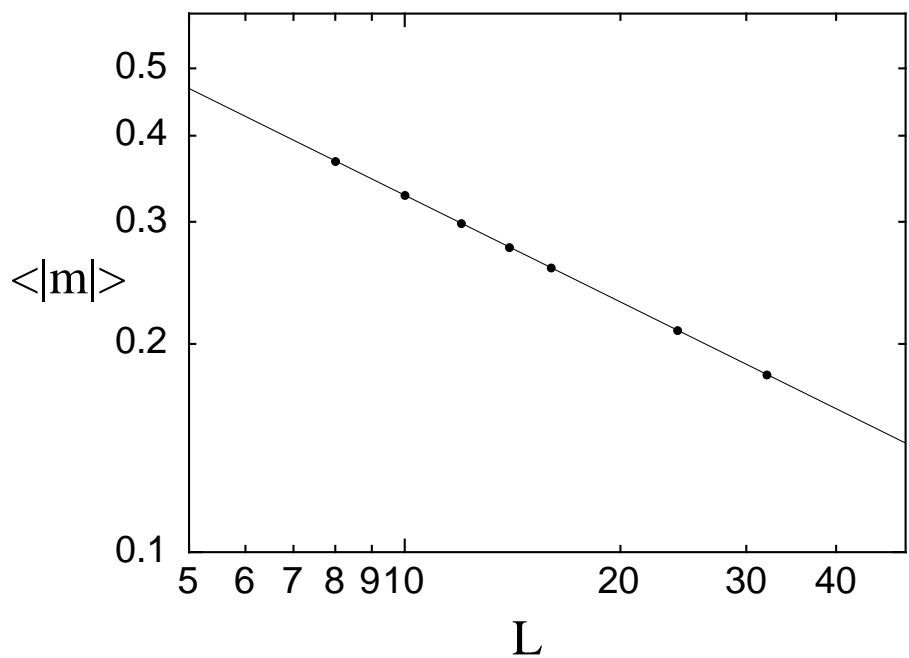}
\caption{
Magnetization $\bra \abs{m}\ket$ at $K_c=0.9360(1)$
as a function of the lattice size $L$.
A least-square fit gives $\beta/\nu=0.5129(7)$.
The jackknife errors for $\bra \abs{m}\ket$ are smaller 
than 1/10 of 
the size of symbols. 
}
\label{fig:5}
\end{figure}

The scaling relation of the magnetization $\bra \abs{m} \ket$ at 
$\Kc$
is given by 
\begin{equation}
\bra \abs{m} \ket_{\Kc} \sim L^{- \beta / \nu}\;.
\label{eqn:scale2}
\end{equation}
We study the scaling of $\bra \abs{m} \ket_{\Kc}$ at $\Kc=0.9360$
and obtain $\beta/{\nu}=0.5129(7)$ from the slope of the fitted 
line 
in Fig.~\ref{fig:5}.  
The fits excluding $L=8$, and $L=8$ and 10 give the results 
consistent with this value 
($\beta/{\nu}$($L=8$ excluded)=0.5130(15) and $\beta/{\nu}$($L=8$
and 10 excluded)=0.5127(21)).
Not like the case of $\nu$ in the previous subsection,
we find that the effect of the error of the $\Kc$ on the estimate 
of $\beta/{\nu}$ is 
larger than the statistical error 0.0007 at $K_c=0.9360$: 
\begin{eqnarray}
 \frac{\beta}{\nu}(\Kc=0.9359)
-\frac{\beta}{\nu}(\Kc=0.9360)&=&0.0009\;,\nonumber \\
 \frac{\beta}{\nu}(\Kc=0.9360)
-\frac{\beta}{\nu}(\Kc=0.9361)&=&0.0011\;.
\end{eqnarray}
Therefore we should use the value 0.0011
for the error of $\beta/\nu$:
\begin{equation}
\frac{\beta}{\nu}=0.5129(11)\;.
\end{equation}
Combined with our estimate for $\nu$, we have 
\begin{equation}
\beta=0.3836(46)\;.
\end{equation}

\subsection{The result for $\displaystyle \frac{\gamma}{\nu}$}
\begin{figure}[p]
\epsfxsize=12cm \epsfbox{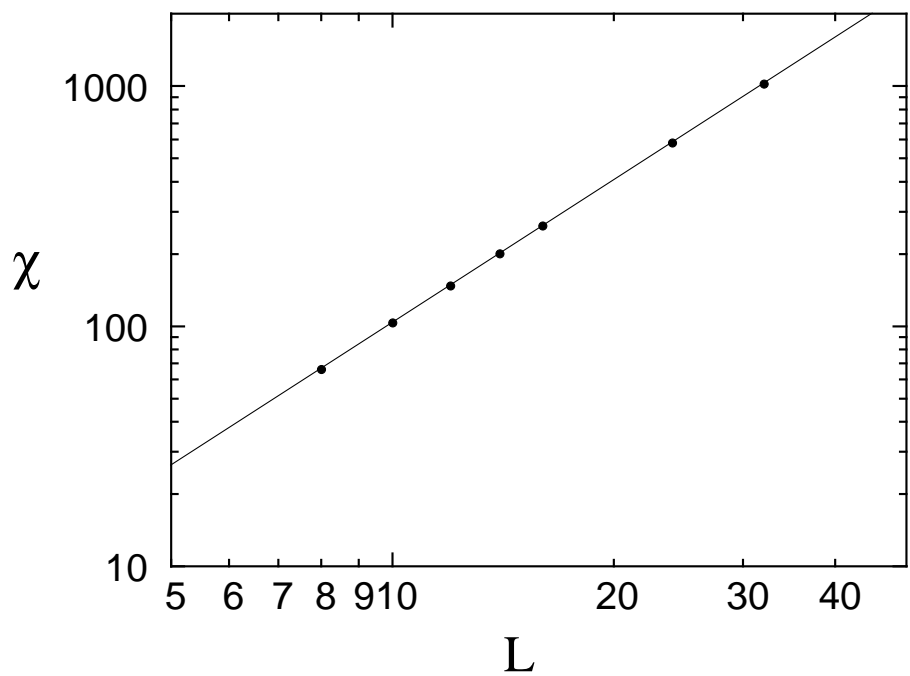}
\caption{
Susceptibility $\chi$ at $K_c=0.9360(1)$ as a function of $L$.
A least-square fit gives $\gamma/\nu=1.9746(15)$.
The jackknife errors for $\chi(K_c)$ are smaller than 1/10 of 
the size of symbols.
}
\label{fig:6}
\end{figure}

For $K\leq \Kc$ the susceptibility $\chi$ is defined by
\begin{equation}
\chi = VK\bra m^2 \ket.
\end{equation}
The scaling relation of $\chi$ at $\Kc$ is given by 
\begin{equation}
\bra \chi \ket _{\Kc} \sim L^{{\gamma}/{\nu}}\;.\label{eqn:scale3}
\end{equation}
With a similar method as in the previous sections, 
we obtain for $\Kc=0.9360$ $\gamma/{\nu}=1.9746(15)$
from the slope of the fitted line in Fig.~\ref{fig:6}.  
Again, the value of $\gamma/{\nu}$ depends strongly on the 
choice of $\Kc$.
\begin{eqnarray}
&& \frac{\gamma }{\nu}(\Kc=0.9360)
  -\frac{\gamma }{\nu}(\Kc=0.9359)\nonumber \\
&\cong& \frac{\gamma }{\nu}(\Kc=0.9361)
    -\frac{\gamma }{\nu}(\Kc=0.9360)\\
&=& 0.0038\;.
\end{eqnarray}
Therefore we quote
\begin{equation}
\frac{\gamma}{\nu}=1.9746(38)\;.
\label{eqn:gammanu}
\end{equation}
Combined with our estimate of $\nu$,
we get 
\begin{equation}
\gamma=1.477(18)\;.
\end{equation}

Using our independent results for ${\beta}/{\nu}$ and 
${\gamma}/{\nu}$,
we can check the hyperscaling relation 
\begin{equation}
\frac{\beta}{\nu}+\frac{1}{2}\;\frac{\gamma}{\nu}-\frac{d}{2}=0\;.
\end{equation}
We find 
\begin{equation}
l.h.s. = 0.0002 \pm 0.003
\end{equation}
that is consistent with zero to $O(10^{-3})$.

\subsection{Scaling of $\chi^c$ and $\Kchi$}
\begin{figure}[p]
\epsfxsize=12cm \epsfbox{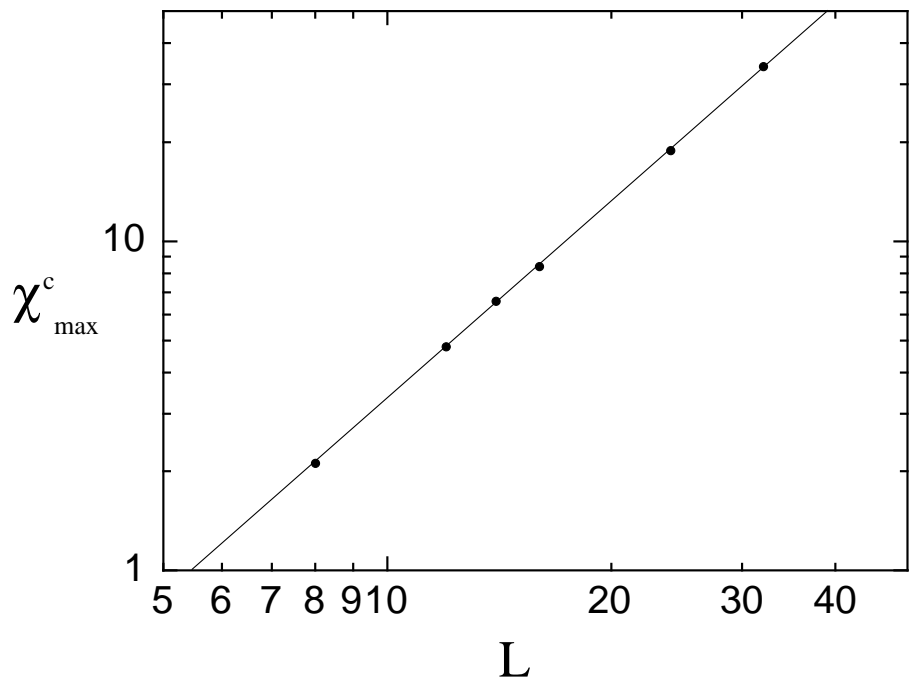}
\caption{
The maximum height of the connected susceptibility $\chi^c_{max}$.
A least-square fit gives $(\gamma/\nu)_c=1.996(8)$.
The jackknife errors for $\chi^c_{max}$ are smaller than 1/5 
of the size 
of symbols.
}
\label{fig:7}
\end{figure}

To make a further check of our results for exponents,
we study the finite size scaling property of the peak of 
the connected 
susceptibility $\chi^c$: 
\begin{equation}
\chi^c = VK\left( \bra m^2 \ket -\bra \abs{m}\ket ^2\right).
\end{equation}
whose maximum value is expected to behave as  
\begin{equation}
\chi^c_{max} \sim L^{({\gamma}/{\nu})_c}.
\end{equation} 
Here we add a suffix $c$ for the exponent to make 
clear the way it is defined.
Because the pseudocritical coupling constant $K_{\chi^c_{max}}$ 
where $\chi^c$ gets its maximum value is found to be slightly 
off the
range of the applicability of the histogram reweighting method 
for the data at $\Kc$ (see the discussion
in section \ref{sec:2}), we carry out new 
simulations at $K \simeq \Kchi$ listed in Table \ref{tbl:1} 
determined
by a preparatory simulation. 
With the histogram method applied to these new data 
we estimate accurate values for $\chi^c_{max}$ and $\Kchi$ 
(see Table \ref{tbl:3}). 
From a least-square fit shown in  Fig.~\ref{fig:7}, we obtain
\begin{equation}
(\gamma / \nu)_c=1.996(8)\;.
\end{equation}
This value is slightly larger than that from the scaling of 
$\chi$, 
1.9746(38), given in (\ref{eqn:gammanu}). 
The same tendency is observed 
for the O(3) Heisenberg model \cite{holm,peczak2}.
Because the quality of the fit for $\chi$ is better than that for 
$\chi^c_{max}$, we quote (\ref{eqn:gammanu}) for the value of 
$\gamma/\nu$.

The scaling property of the pseudocritical coupling $\Kchi$ 
provides us 
another test of our results: 
\begin{equation}
\Kchi ^{-1} \sim {\Kc}^{-1} + a L^{-1/\nu}\;.
\end{equation}
Using our estimate $1/{\nu}=1.337$,
we fit the data with two parameters, ${\Kc}^{-1}$ and $a$, 
to obtain
\begin{equation}
K_c=0.9360(2).
\end{equation}
This value is consistent with our $K_c$ from 
the crossing points of the Binder cumulant.

\subsection{Restriction of the transition point by Q value}
The scaling relations (\ref{eqn:scale1}), (\ref{eqn:scale2}),
and (\ref{eqn:scale3}) require that
the estimated value of $\Kc$ is close enough
to the real transition point.
If we fix $\Kc$ far from the real transition point in these 
scaling 
relations the data will not fit them well any more.

The quality of a least-square fit is determined by 
the Q value\cite{press}:
\begin{equation}
Q(\chi^2,n)=\int_{\chi^2}^{\infty}dt{\left(
              \frac{t}{2}\right)}^{\frac{n}{2}-1}
              \e ^{-t/2}
\end{equation}
where $\chi^2$ 
is the weighted sum of squared deviations of data from the fit, 
and 
$n = (\mbox{number of data points})$ $-$ $(\mbox{number of fit 
parameters})$ 
is the degree of freedom for the fit.  
We may consider that the fitting procedure is appropriate 
if $0.1\leq Q \leq 0.9$. 
If, on the other hand, $Q <0.1$ something is wrong: the error of 
data may be under-estimated or the fitting function may be 
incorrect, and 
if $Q > 0.9$ error of data may be over-estimated or we have too 
many 
fit parameters.

\begin{figure}[p]
\epsfxsize=12cm \epsfbox{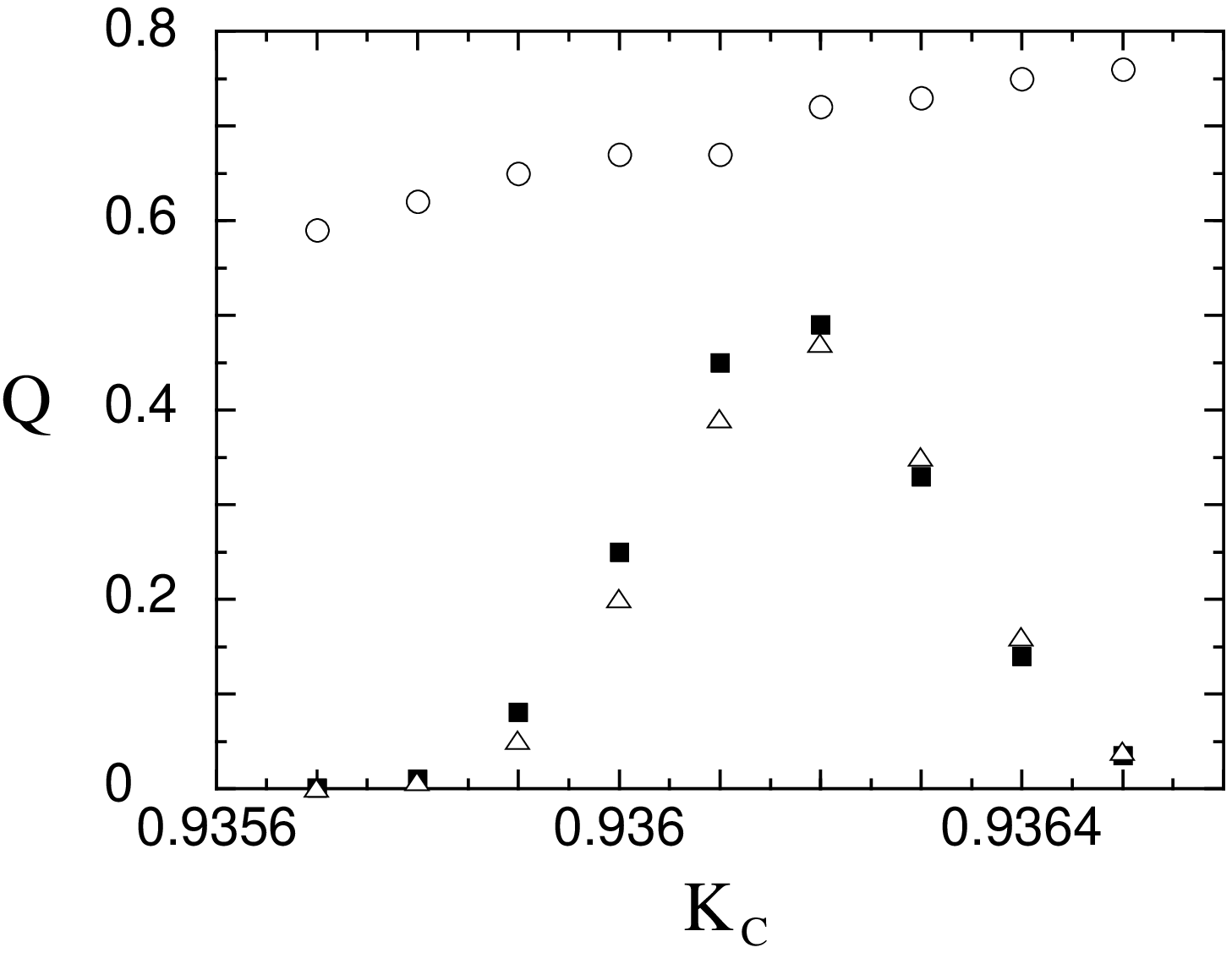}
\caption{
Q values of least-square fits for $dU_L/dK$ (circles), 
$\bra \abs{m}\ket$ (squares), and $\chi$ (triangles) for various 
fixed $\Kc$.
}
\label{fig:8}
\end{figure}

In the present case, if we fix $\Kc$ far from the real 
transition point, 
the quality of the scaling fits must become low so that 
the Q-value decreases to a value less than 0.1.
In Fig \ref{fig:8} the Q-values of our finite-size scaling 
fits for
$\displaystyle \dudk$, $\bra \abs{m}\ket$ and $\chi$ are plotted 
as a function of $\Kc$.
We find that Q-value for $\displaystyle \dudk$ is not so 
sensitive on $\Kc$, 
while the Q-values for $\bra \abs{m}\ket$ and $\chi$ depend 
sensitively on
$\Kc$.
This difference of the dependence on $\Kc$ between 
$\displaystyle \dudk$ and 
$\bra \abs{m}\ket$, $\chi$ is the same as that observed 
for the O(3) Heisenberg model \cite{holm}.    
From the condition that $Q \le 0.1$  we have
\begin{equation}
0.9359 \leq \Kc \leq 0.9364.
\end{equation}
This provides us another consistency check of our analyses. 
The value obtained from the crossing point of the Binder cumulant
$\Kc=0.9360(1)$ is well included in this region.

\subsection{Comparison with the results of the $4-\epsilon$ 
expansion} 
The critical exponents obtained in this work 
are compiled in Table \ref{tbl:2}  together with
the values by the $4-\epsilon$ expansion method\cite{baker}. 
In our results, the exponents $\gamma/\nu$, $\beta/\nu$, and 
$\nu$ 
are determined independently and $\alpha$ and $\delta$ are 
calculated 
using (hyper)scaling relations with the value of other exponents.
In the results of the $4-\epsilon$ expansion method, 
$\eta $ and $ \nu$ are estimated independently
and other exponents are calculated with $\eta $ and $ \nu$.
Our results are completely consistent with those of 
$4-\epsilon$ with
the errors reduced to about halves of them.

\section{Conclusion}
\label{sec:4}

We simulated the three-dimensional O(4) Heisenberg model
by applying the single-cluster algorithm,
which reduces the dynamical exponent to about zero.
The histogram reweighting method with high statistics data 
confirmed that the transition is second order for this model.
We performed a precise estimation of the 
critical point to get $\Kc = 0.9360(1)$ 
from the crossing point of the Binder cumulant.
The critical exponents were calculated using finite-size scaling 
at $\Kc$.
The exponents obtained, 
which are summarized in Table \ref{tbl:2}, are completely 
consistent with 
those of the $4-\epsilon$ method with 
the errors reduced to about halves of them.

We are grateful to Y.\ Iwasaki and A.\ Ukawa 
for valuable discussions and helpful suggestions. 
Numerical calculations were done on HITAC S820/80 at the 
National Laboratory for High Energy Physics (KEK). We would 
like to thank members of KEK for their hospitality and strong 
support, and 
members of the Theory Group of the Institute of Physics, 
University of 
Tsukuba, for innumerable discussions and encouragements. 
This project is in part supported by a Grant-in-Aid of 
Ministry of 
Education, Science and Culture (No.6206001 and 02402003).

\clearpage
\begin{table}[h]
\begin{tabular}{|c||c|c|c|c|c|c|}
\hline
$L$ & simulation point ($K$)& sweeps/$10^3$ & $\bra C \ket$
&$\tau_m$ &$\tau_c$ &${\rm sweeps}/(\tau_c \times 10^3)$\\
\hline
\hline
8  &  0.892  &  3000  &  31.6 & 1.95  & 31.6 &  95 \\ 
\hline
8  &  0.935  &  3000  &  51.1 & 2.49  & 24.9 &  120 \\
\hline
\hline
10 &  0.935  &  4400  &  78.7 & 2.55  & 32.4 &  136 \\
\hline
\hline
12 &  0.910  &  4000  &  67.3 & 1.98  & 49.8 &   80 \\
\hline
12 &  0.935  &  6000  &  113.2& 2.48  & 37.7 &   159 \\
\hline
\hline
14 &  0.912  &  1400  &  88.0 & 1.82  & 56.6 &   25  \\
\hline
14 &  0.935  &  6000  &  154.5& 2.47  & 43.9 &   136  \\
\hline
\hline
16 &  0.920  &  3000  &  129.5& 1.98  & 62.2 &   23  \\
\hline
16 &  0.935  &  5200  &  197.4& 2.32  & 48.1 &   108  \\
\hline
\hline
24 &  0.926  &  1500  &  280.9& 1.87  & 92.0 &   16   \\
\hline
24 &  0.935  &  2400  &  427.1& 2.21  & 71.4 &   34   \\
\hline
\hline
32 &  0.928  &  1500  &  449.5& 1.62  & 117.9 &   13   \\
\hline
32 &  0.935  &  5600  &  719.1& 1.95  & 88.9  &   63  \\
\hline
\end{tabular}
\caption{Simulation parameters and statistics. 
        $\bra C \ket$ is the mean cluster volume; 
        $\tau_c$ is the autocorrelation time in units of sweeps, 
	i.e.\ in units of the number of cluster updates; 
        $\tau_m$ is the autocorrelation time converted into 
Metropolis units, 
	i.e.\ in units of updates of whole spins on the lattice: 
	$\displaystyle \tau_m=\tau_c \bra C \ket / V$, 
	where $V=L^3$ is the lattice volume.
\label{tbl:1}}
\end{table}

\clearpage

\begin{table}[h]
\begin{tabular}{|c|c|c|}
\hline
\hline
&\makebox[4.5cm]{$4-\epsilon$} &\makebox[4.5cm]{This study} \\
\hline
$ \gamma / \nu=2-\eta$ & 1.97(1)     & 1.9746(38)  \\
$\beta / \nu$         & 0.515(5)    & 0.5129(11)  \\
$\nu$                         & 0.73(2)     & 0.7479(90) \\
$\gamma$                      & 1.44(4)     & 1.477(18)   \\
$\beta$                      & 0.38(1)      & 0.3836(46)  \\
$\delta$                      & 4.82(5)     & 4.851(22)  \\
$\alpha=2-d\nu$              & $-0.19(6)$     & $-0.244(27)$ \\
\hline
\hline
\end{tabular}
\protect\caption{Critical exponents 
of the three-dimensional O(4) Heisenberg model
obtained by a study with the $4-\epsilon$ expansion method
and by this study. The calculation with the $4-\epsilon$ 
expansion method 
is done to seven loops \protect\cite{baker} and the results 
are quoted in Ref.~\protect\cite{wilczek2}. 
In the $4-\epsilon$ method, independent calculations are done for 
$\nu$ and $\eta$.
In this study, $\gamma/\nu$, $\beta/\nu$ and $\nu$ are determined 
independently. 
Other exponents are calculated using (hyper)scaling relations.
\label{tbl:2}}
\end{table}

\clearpage

\begin{table}[h]
\begin{tabular}{|c|c|c|c|}
\hline
\hline
\makebox[1cm]{$L$}&\makebox[2.5cm]{$K_{simu}$}&
\makebox[2.5cm]{$\Kchi$}&\makebox[2.5cm]{$\chi^c_{max}$}\\
\hline
8 & 0.892 & 0.8907(37) & 2.118(09)\\ 
12& 0.910 & 0.9109(05) & 4.793(28)\\ 
14& 0.912 & 0.9144(06) & 6.577(48)\\ 
16& 0.920 & 0.9183(08) & 8.390(46)\\ 
24& 0.926 & 0.9253(06) & 18.89(16)\\
32& 0.928 & 0.9289(01) & 33.97(32)\\
\hline
\hline
\end{tabular}
\caption{Results for the pseudocritical coupling $\Kchi$ and 
the maximum values $\chi^c_{max}$ of the connected susceptibility
 $\chi^c$. 
The $K$ dependence of $\chi^c$ is determined 
by the histogram reweighting method using the data simulated at 
$K_{simu}$  
on an $L^3$ lattice. \label{tbl:3}}
\end{table}

\clearpage

\end{document}